\newcommand{\CLASSINPUTtoptextmargin}{0.75in}%
\newcommand{\CLASSINPUTbottomtextmargin}{1in}%
\newcommand{\CLASSINPUTinnersidemargin}{0.63in}%
\newcommand{\CLASSINPUToutersidemargin}{0.63in}%
\documentclass[conference,10pt,letterpaper]{IEEEtran}%
\usepackage{amsmath}
\usepackage{graphicx}
\usepackage{multirow}
\usepackage[none]{hyphenat}
\usepackage{float}
\usepackage{subfig}
\usepackage{dblfloatfix}
\usepackage{t1enc}
\usepackage{times}
\makeatletter

\def\@IMSauthorblockNAMEstyle{\normalfont\IMSauthorsize}
\def\@IMSauthorblockAFFILstyle{\normalfont\IMSaffilsize}
\def\@IMSauthorblockEMAILstyle{\normalfont\IMSaffilsize}
\def\IMSauthorblockNAME#1{%
\relax\@IMSauthorblockNAMEstyle%
#1%
}%
\def\IMSauthorblockAFFIL#1{%
\relax\@IMSauthorblockAFFILstyle%
\vskip\@IEEEauthorblockAtopspace
#1%
}%
\def\IMSauthorblockEMAIL#1{%
\relax\@IMSauthorblockEMAILstyle%
\vskip\@IEEEauthorblockAtopspace
#1%
}%
\newcommand{\IMSauthor}[1]{%
\ifIsBlindReviewVersion%
\author{\phantom{\parbox{\textwidth}{\center\relax#1}}}%
\else%
\author{\parbox{\textwidth}{\center\relax#1}}%
\fi%
}%
\newif\ifIsBlindReviewVersion

\def\IMSthispaperforfinalpublication{\IsBlindReviewVersionfalse}
\def\@maketitle{\newpage
\bgroup\par\addvspace{0.5\baselineskip}\centering%
\ifCLASSOPTIONtechnote
   {\bfseries\large\@IEEEcompsoconly{\sffamily}\@title\par}\vskip 1.3em{\lineskip .5em\@IEEEcompsoconly{\sffamily}\@author
   \@IEEEspecialpapernotice\par{\@IEEEcompsoconly{\vskip 1.5em\relax
   \@IEEEtitleabstractindextextbox{\@IEEEtitleabstractindextext}\par
   \hfill\@IEEEcompsocdiamondline\hfill\hbox{}\par}}}\relax
\else
   \vskip0.2em{\IMStitlesize\ifCLASSOPTIONtransmag\bfseries\LARGE\fi\@IEEEcompsoconly{\sffamily}\@IEEEcompsocconfonly{\normalfont\normalsize\vskip 2\@IEEEnormalsizeunitybaselineskip
   \bfseries\Large}\@title\par}\vskip1.0em\par
   \ifCLASSOPTIONconference%
      {\@IEEEspecialpapernotice\mbox{}\vskip\@IEEEauthorblockconfadjspace%
       \mbox{}\hfill\begin{@IEEEauthorhalign}\@author\end{@IEEEauthorhalign}\hfill\mbox{}\par}\relax
   \else
      \ifCLASSOPTIONpeerreviewca
         {\@IEEEcompsoconly{\sffamily}\@IEEEspecialpapernotice\mbox{}\vskip\@IEEEauthorblockconfadjspace%
          \mbox{}\hfill\begin{@IEEEauthorhalign}\@author\end{@IEEEauthorhalign}\hfill\mbox{}\par
          {\@IEEEcompsoconly{\vskip 1.5em\relax
           \@IEEEtitleabstractindextextbox{\@IEEEtitleabstractindextext}\par\hfill
           \@IEEEcompsocdiamondline\hfill\hbox{}\par}}}\relax
      \else
         \ifCLASSOPTIONtransmag
           {\@IEEEspecialpapernotice\mbox{}\vskip\@IEEEauthorblockconfadjspace%
            \mbox{}\hfill\begin{@IEEEauthorhalign}\@author\end{@IEEEauthorhalign}\hfill\mbox{}\par
           {\vspace{0.5\baselineskip}\relax\@IEEEtitleabstractindextextbox{\@IEEEtitleabstractindextext}\vspace{-1\baselineskip}\par}}\relax
         \else
           {\lineskip.5em\@IEEEcompsoconly{\sffamily}\sublargesize\@author\@IEEEspecialpapernotice\par
           {\@IEEEcompsoconly{\vskip 1.5em\relax
            \@IEEEtitleabstractindextextbox{\@IEEEtitleabstractindextext}\par\hfill
            \@IEEEcompsocdiamondline\hfill\hbox{}\par}}}\relax
         \fi
      \fi
   \fi
\fi\par\addvspace{0.0\baselineskip}\egroup}

\def\IMStitlesize{\@setfontsize{\IMStitlesize}{18}{21pt}}
\def\IMSauthorsize{\@setfontsize{\IMSauthorsize}{12}{13pt}}
\def\IMSaffilsize{\@setfontsize{\IMSaffilsize}{12}{13pt}}
\def\IMScaptionsize{\@setfontsize{\IMScaptionsize}{8}{9pt}}
\def\IMSbibsize{\@setfontsize{\IMSbibsize}{8}{9pt}}

\def\@IEEEauthorblockNstyle{\IMSauthorsize\@IEEEcompsocnotconfonly{\sffamily}\@IEEEcompsocconfonly{\large}}
\def\@IEEEauthorblockAstyle{\IMSaffilsize\@IEEEcompsocnotconfonly{\sffamily}\@IEEEcompsocconfonly{\itshape}\@IEEEcompsocconfonly{\large}}
\def\@IEEEauthordefaulttextstyle{\IMSauthorsize\@IEEEcompsocnotconfonly{\sffamily}\sublargesize}

\def\thebibliography#1{\section*{\refname}%
    \addcontentsline{toc}{section}{\refname}%
    \IMSbibsize\@IEEEcompsocconfonly{\small}\vskip 0.3\baselineskip plus 0.1\baselineskip minus 0.1\baselineskip
    \list{\@biblabel{\@arabic\c@enumiv}}%
    {\settowidth\labelwidth{\@biblabel{#1}}%
    \leftmargin\labelwidth
    \advance\leftmargin\labelsep\relax
    \itemsep \IEEEbibitemsep\relax
    \usecounter{enumiv}%
    \let\p@enumiv\@empty
    \renewcommand\theenumiv{\@arabic\c@enumiv}}%
    \let\@IEEElatexbibitem\bibitem%
    \def\bibitem{\@IEEEbibitemprefix\@IEEElatexbibitem}%
\def\newblock{\hskip .11em plus .33em minus .07em}%
\ifCLASSOPTIONtechnote\sloppy\clubpenalty4000\widowpenalty4000\interlinepenalty100%
\else\sloppy\clubpenalty4000\widowpenalty4000\interlinepenalty500\fi%
    \sfcode`\.=1000\relax}

\long\def\@makecaption#1#2{%
\ifx\@captype\@IEEEtablestring%
\par\@IEEEtabletopskipstrut
\else
\@IEEEfigurecaptionsepspace
\fi
\setbox\@tempboxa\hbox{\normalfont\IMScaptionsize {#1.}\nobreakspace\nobreakspace #2}%
\ifdim \wd\@tempboxa >\hsize%
\setbox\@tempboxa\hbox{\normalfont\IMScaptionsize {#1.}\nobreakspace\nobreakspace}%
\parbox[t]{\hsize}{\normalfont\IMScaptionsize\noindent\unhbox\@tempboxa#2}%
\else
\ifCLASSOPTIONconference \hbox to\hsize{\normalfont\IMScaptionsize\hfil\box\@tempboxa\hfil}%
\else \hbox to\hsize{\normalfont\IMScaptionsize\box\@tempboxa\hfil}%
\fi\fi
\ifx\@captype\@IEEEtablestring%
\@IEEEtablecaptionsepspace
\else
\fi}

\newlength\tablecaptiontotableskip
\newlength\figuretocaptionskip
\def\@IEEEfigurecaptionsepspace{\vskip\figuretocaptionskip\relax}%
\def\@IEEEtablecaptionsepspace{\vskip\tablecaptiontotableskip\relax}%

\def\abstract{\normalfont%
\@IEEEabskeysecsize\bfseries\textit{\abstractname}\,\bfseries\textit{---}\,%
\@IEEEgobbleleadPARNLSP}%

\def\IEEEkeywords{\normalfont%
\@IEEEabskeysecsize\bfseries\textit{\IEEEkeywordsname}\,\bfseries\textit{---}\,%
\@IEEEgobbleleadPARNLSP}%
\def\endIEEEkeywords{\relax\vspace{0.67ex}%
\par\if@twocolumn\else\endquotation\fi%
\normalsize\normalfont}%

\def\@IEEEauthorblockNtopspace{0ex}
\def\@IEEEauthorblockAtopspace{1mm}
\def\tablename{Table}
\def\thetable{\arabic{table}}
\def\IEEEkeywordsname{Keywords}
\def\subsubsection{\@startsection{subsubsection}{3}{\z@}{1.5ex plus 1.5ex minus 0.5ex}%
{0.7ex plus .5ex minus 0ex}{\normalfont\normalsize\itshape}}%
\def\@seccntformat#1{\csname the#1dis\endcsname\relax}
\def\thesectiondis{\thesection.\hskip 0.5em}
\def\thesubsectiondis{{\hbox to\parindent{\Alph{subsection}.}}}
\def\thesubsubsectiondis{{\hbox to \parindent{\arabic{subsubsection})}}}
\def\theparagraphdis{{\hbox to \parindent{\alph{paragraph})}}}
\IEEEilabelindentA \parindent
\IEEEilabelindent \IEEEilabelindentA
\IEEEelabelindent \parindent
\IEEEdlabelindent \parindent
\IEEElabelindent \parindent

\newlength\@IMSparindent
\newcommand\IMSdisplayacksection[1]{%
\ifIsBlindReviewVersion%
\noindent\phantom{\parbox[t]{\columnwidth}{\normalbaselines\setlength{\parindent}{\@IMSparindent}{#1}\strut}}
\else%
\noindent\parbox[t]{\columnwidth}{\normalbaselines\setlength{\parindent}{\@IMSparindent}{#1}\strut}%
\fi%
}%

\makeatother

\graphicspath{{./}} 
\begin{document}
\raggedbottom
%
%
%
\title{Efficient and Accurate Method for Separating Variant Components from Invariant Background and Component Model Fusion for \\Fast RFIC Design Space Exploration}
%
%
%
\IMSthispaperforfinalpublication
\IMSauthor{%
\IMSauthorblockNAME{
Hongyang Liu and Dan Jiao
}
\\%
\IMSauthorblockAFFIL{
School of Electrical and Computer Engineering, Purdue University, West Lafayette, IN 47907, USA
}
\\%
\IMSauthorblockEMAIL{
liu3572@purdue.edu, 
djiao@purdue.edu
}
}
%
\maketitle
%

\renewcommand{\thefootnote}{} 
\footnotetext{This work was supported by Rapid-HI (Heterogeneous Integration) Design
Institute (an Elmore ECE Emerging Frontiers Center), a grant from NSF
under award CCF-2235414, a grant from SRC under award 2878.033, and a DARPA NGMM grant.

\vspace{0.5em} \noindent \copyright~2026 IEEE. Personal use of this material is permitted.  Permission from IEEE must be obtained for all other uses, in any current or future media, including reprinting/republishing this material for advertising or promotional purposes, creating new collective works, for resale or redistribution to servers or lists, or reuse of any copyrighted component of this work in other works.}
%
%
\begin{abstract}
The design of RFIC often involves exploring a large number of design variations in an invariant background composed of the processing stack and unchanged circuit blocks. Conventional electromagnetic solvers require a full-domain simulation for every design variation. In this work, we present a fast method that effectively separates the variant components from the  invariant background. It algebraically decomposes the total field solution into the contributions from the design-dependent variations and the invariant background. Hence, the field response due to the invariant background can be simulated once and reused for all design variations. Only the variant components need to be simulated at each design variation, the size of which is small. We also develop an efficient way of reusing the model of each component and fusing them accurately to obtain the model of a system composed of many components. The reduced system of variant components involves computing the field solutions in the invariant background due to all possible sources located at variant components, the number of which can be large. We develop a fast algorithm to reduce them to a few field solutions, the number of which is on the order of the layer number.  The proposed method has been applied to RFIC design space exploration. Its accuracy, robustness, and efficiency have been demonstrated. 
\end{abstract}
%

\section{Introduction} \label{sec:introduction}
\IEEEPARstart{T}{he} design and optimization of Radio Frequency Integrated Circuit (RFIC) components and systems often involve the exploration of different components, their placement, topology, geometrical parameters, materials, etc., in a large background that does not change \cite{Razavi:2011:RFMicro, Cadence:2022:EMAnalysis}. The invariant background consists of the entire set of materials and physical structures that do not change in the design considerations. For example, it can be the layered dielectric stack for a given processing technology; it can also be the layered stack in conjunction with the circuit blocks that are not varied in the design. The design space to explore is also generally large, with a combinatorial number of choices that can be prohibitive. Conventional partial differential equation (PDE)-based full-wave simulation methods do not separate the variant components from the invariant parts. This results in a full-domain simulation for every design variation, which is computationally expensive. Domain decomposition (DD)-based methods typically divide a computational domain into non-overlapping or overlapping subdomains, whose union constitutes the entire domain \cite{Lu:2025:EMISimDDM, Zhang:2021:ScalableDDM, Zhao:2003:InductanceDDM, He:2006:DDM-EA, DeLaRubia:2007:MicrowaveDDM, Kleis:1994:DDMCircuitSim}. However, the invariant background itself occupies the entire computational domain. Its separation from the variant components, which are located anywhere in the background, cannot be addressed by a conventional DD technique. 

Moreover, the components in an RFIC couple with each other via the common background, especially at higher frequencies \cite{Mandal:2006:PackagingEffects, Sudo:2004:EMISOP, Aragones:2003:PackageEffects, Nagata:2012:SubstrateNoise, Poon:2009:QuadrupoleInductors, Banerjee:2001:OnChipInductance, Martin:2016:TwistedInductor, Chen:2022:CouplingSuppression}. When a component is designed in an RFIC, designers would like to reuse the model built for the component to design a system composed of many components. However, due to the electromagnetic coupling between the RFIC components, there is no obvious way to reuse the individual model of each component to obtain the model for multiple components placed and connected in an arbitrary manner in the physical layout. For example, an inductor model is built for one processing technology. However, when five of these inductors are placed in the physical layout of the same processing technology, the resultant circuit performance cannot be determined from the simple stitching of individual models. How to reuse the model of each component to build the model of a system composed of multiple components while accurately capturing the coupling among them remains a research problem. The same problem exists in an AI-based design when model fusion is required to build an AI system model \cite{Li:2023:AIModelFusion}.

This work is developed to address the aforementioned research problems. Our method effectively separates variant components from the invariant background to simulate, and with first-principles accuracy. It algebraically decomposes the total electromagnetic field into contributions from the invariant background and the varying components. The former only needs to be computed once and reused for all design variations. The latter involves the simulation of a small system containing only the variant components. For each design variation, only the latter needs to be computed. Furthermore, we show how to efficiently reuse the model obtained for each component to create the global model for the RFIC system composed of many components. Moreover, we develop a fast technique to reduce the computation of field solutions due to all possible source locations in the invariant background, i.e., the entire numerical Green's function matrix in the background, to the computation of a few field solutions, the number of which is on the order of the layer number. The proposed method has been applied to state-of-the-art RFIC designs and has demonstrated superior performance in accuracy and efficiency.
\section{Proposed Method} \label{sec:separation}
The electromagnetic phenomena in RFIC components and systems from DC to high frequencies are governed by Maxwell’s equations, which can be discretized into the following linear system of equations in the frequency domain:
\begin{equation}
\label{equ:Maxwell_Freq_LinearSystem}
\left[-\omega^2 \overline{\mathbf{D}}_\epsilon(\mathbf{p})+j \omega \overline{\mathbf{D}}_\sigma(\mathbf{p})+\overline{\mathbf{S}}(\mathbf{p})\right] \mathbf{e}(\mathbf{p})=-j \omega \mathbf{J}(\mathbf{p}), 
\end{equation}
where $\mathbf{e}$ denotes the unknown electric field vector to be solved for, $\mathbf{J}$ is a vector of current density, $\overline{\mathbf{D}}_{\epsilon}$ and $\overline{\mathbf{D}}_{\sigma}$ are matrices related to permittivity and conductivity, respectively, $\overline{\mathbf{S}}$ denotes the discretized $\nabla\times\frac{1}{\mu}\nabla\times$  operator with $\mu$ being permeability, and $\mathbf{p}$ is a vector representing design parameters and variations of an RFIC. Here, all matrices are sparse in a PDE-based solution. In a finite-difference based solution, $\overline{\mathbf{D}}_{\epsilon}$ and $\overline{\mathbf{D}}_{\sigma}$ are diagonal matrices of permittivity and conductivity.

The invariant background consists of all materials and physical structures that do not change in the design variation. Let its corresponding system matrix be denoted by $\overline{\mathbf{Y}}_b$; hence, we have 
\begin{equation} 
\label{equ:Y0System} 
\overline{\mathbf{Y}}_b =-\omega^2 \overline{\mathbf{D}}_{\epsilon, b}+j \omega \overline{\mathbf{D}}_{\sigma, b}+\overline{\mathbf{S}}_b. 
\end{equation}
When the variant components are present, the full system matrix $\overline{\mathbf{Y}}(\mathbf{p})$ from (\ref{equ:Maxwell_Freq_LinearSystem}) can be expressed as an update of the background system as follows:
\begin{equation} 
\label{equ:YSystem} 
\overline{\mathbf{Y}}(\mathbf{p})=\overline{\mathbf{Y}}_b+ \overline{\mathbf{Y}}_v(\mathbf{p}), 
\end{equation}
in which $\overline{\mathbf{Y}}_v$  arises from the variant components. Since most of the circuit design focuses on the conductive parts, $\overline{\mathbf{Y}}_v$ can be written as
\begin{equation} 
\label{equ:DeltaDSigma} 
\overline{\mathbf{Y}}_v(\mathbf{p})=j \omega \Delta \overline{\mathbf{D}}_\sigma(\mathbf{p})=j \omega[\overline{\mathbf{D}}_\sigma(\mathbf{p})-\overline{\mathbf{D}}_{\sigma,b}]. 
\end{equation}
If the layered stack is treated as the background, then the above is simply a diagonal matrix of conductivity at the circuit component being designed. Note that the proposed method remains applicable to the case in which the permittivity change is also considered in the design variation. The dimension of $\overline{\mathbf{Y}}$ is $N$, but the variant components occupy only a small portion of the physical layout. Hence, \eqref{equ:DeltaDSigma}  is of low rank, which can be written as: 
\begin{equation}
\label{equ:lowRankUpdate}
\overline{\mathbf{Y}}_v(\mathbf{p})=\overline{\mathbf{I}}_v(\mathbf{p})_{N\times k} \overline{\mathbf{D}}_v(\mathbf{p})_{k\times k} \overline{\mathbf{I}}_v(\mathbf{p})^T_{k\times N}, 
\end{equation}
where $\overline{\mathbf{D}}_v(\mathbf{p})$ is a $k \times k$ diagonal matrix whose entries are $j \omega \Delta \sigma $ at the unknowns corresponding to variant components, and $\overline{\mathbf{I}}_v(\mathbf{p})$ is an $N \times k$ identity matrix composed of $k$ unit vectors, each of which has $1$ appearing at one of the variant component unknowns. 

With this low-rank form, we can apply the Sherman–Morrison–Woodbury formula \cite{Sherman:1950:Adjustment} to the inverse of the full system matrix from (\ref{equ:YSystem}), obtaining 
\begin{equation}
\label{equ:WoodburyESol}
\begin{aligned}
&\mathbf{e}(\mathbf{p}) = \mathbf{e}_b(\mathbf{p})- \\
& \overline{\mathbf{Y}}_b^{-1} \overline{\mathbf{I}}_v(\mathbf{p}) \left[\overline{\mathbf{D}}_v(\mathbf{p})^{-1}+\overline{\mathbf{I}}_v(\mathbf{p})^T \overline{\mathbf{Y}}_b^{-1} \overline{\mathbf{I}}_v(\mathbf{p})\right]^{-1}_{k\times k}\overline{\mathbf{I}}_v(\mathbf{p})^T\mathbf{e}_b(\mathbf{p}),
\end{aligned}
\end{equation}
in which 
\begin{equation}\label{equ:eb}
  \mathbf{e}_b(\mathbf{p})=-j\omega\overline{\mathbf{Y}}_b^{-1}\mathbf{J}(\mathbf{p}).
\end{equation}

Equation \eqref{equ:WoodburyESol} has a clear physical meaning. The matrix $\overline{\mathbf{Y}}_b^{-1}$ represents the complete $N \times N$ numerical Green's function of the invariant background, whereas $\overline{\mathbf{Y}}_b^{-1} \overline{\mathbf{I}}_v(\mathbf{p})$ contains its $k$ columns due to the $k$ sources located at the variant components. Let $\overline{\mathbf{G}}_k(\mathbf{p})=\overline{\mathbf{Y}}_b^{-1} \overline{\mathbf{I}}_v(\mathbf{p})$. The $k \times k$ dense matrix to be inverted, denoted by $\overline{\mathbf{C}}(\mathbf{p})$,  
\begin{equation}
\label{equ:MExpression}
\overline{\mathbf{C}}(\mathbf{p}) = \overline{\mathbf{D}}_v(\mathbf{p})^{-1}+\overline{\mathbf{I}}_v(\mathbf{p})^T \overline{\mathbf{G}}_k(\mathbf{p})
\end{equation}
describes how the variant components are coupled. As can be seen from \eqref{equ:WoodburyESol}, we algebraically decompose the total field $\mathbf{e}(\mathbf{p})$ into the contributions from the invariant background and the variant components. The former is the first component shown on the right hand side of \eqref{equ:WoodburyESol}, which is \eqref{equ:eb}. The latter is the second component, found by solving the $k$ equivalent source vectors, $-\overline{\mathbf{C}}(\mathbf{p})^{-1} \overline{\mathbf{I}}_v(\mathbf{p})^T\mathbf{e}_b(\mathbf{p})$, induced on the variant components, and then using them to radiate to the background via the background Green's function, manifested by the front multiplication with $\overline{\mathbf{G}}_k(\mathbf{p})$.

Using the aforementioned approach, we effectively separate the large invariant background from small variant components to facilitate rapid design space exploration. The field solution due to the invariant background for any source, shown in \eqref{equ:eb}, only needs to be evaluated once and then reused for all design variations. The simulation of each design variation involves only the variant components, the size of which is $k<<N$. Specifically, it is the construction of the small coupled system \eqref{equ:MExpression} and its solution for a few right-hand sides. Moreover, the proposed approach also produces an efficient and accurate way for component fusion. To elaborate, when one component is simulated in the invariant background of an RFIC design, we essentially obtain $\overline{\mathbf{Y}}_b^{-1} \overline{\mathbf{I}}_{v,i}$, where $\overline{\mathbf{I}}_{v,i}$ represents the unit source vectors located in the $i$-th variant component. The model of \eqref{equ:WoodburyESol} shows that we can reuse the $\overline{\mathbf{Y}}_b^{-1} \overline{\mathbf{I}}_{v,i}$ obtained for an arbitrary $i$-th component to find the solution for a system of many components. Taking a 3-component system as an example, we combine each component's model as $[\overline{\mathbf{Y}}_b^{-1} \overline{\mathbf{I}}_{v,1} \ \overline{\mathbf{Y}}_b^{-1}\overline{\mathbf{I}}_{v,2} \ \overline{\mathbf{Y}}_b^{-1} \overline{\mathbf{I}}_{v,3}]_{N\times(k_1+k_2+k_3)}$, and then select the rows, i.e., the field solutions, at the variant components. By adding the resultant, which is a square matrix of size $k_1+k_2+k_3$, to the diagonal matrix of $\overline{\mathbf{D}}_v(\mathbf{p})^{-1}$ at the corresponding components and solving the resulting small matrix, we obtain the result for the coupled system.
\section{Efficient Computation} \label{sec:effientGF}
Equation \eqref{equ:WoodburyESol} shows that the response from the variant components depends on the Green's function generated by unit sources located at the variant components, i.e., $\overline{\mathbf{G}}_k(\mathbf{p})=\overline{\mathbf{Y}}_b^{-1} \overline{\mathbf{I}}_v(\mathbf{p})$. A conventional method to compute it would require solving the background system matrix $\overline{\mathbf{Y}}_b$ for all $k$ right-hand sides contained in $\overline{\mathbf{I}}_v(\mathbf{p})$. However, $k$ can be on the order of tens of thousands in a large-scale design space exploration, making the computation inefficient. 

\begin{figure}[!t]
\centering
\includegraphics[width=\columnwidth]{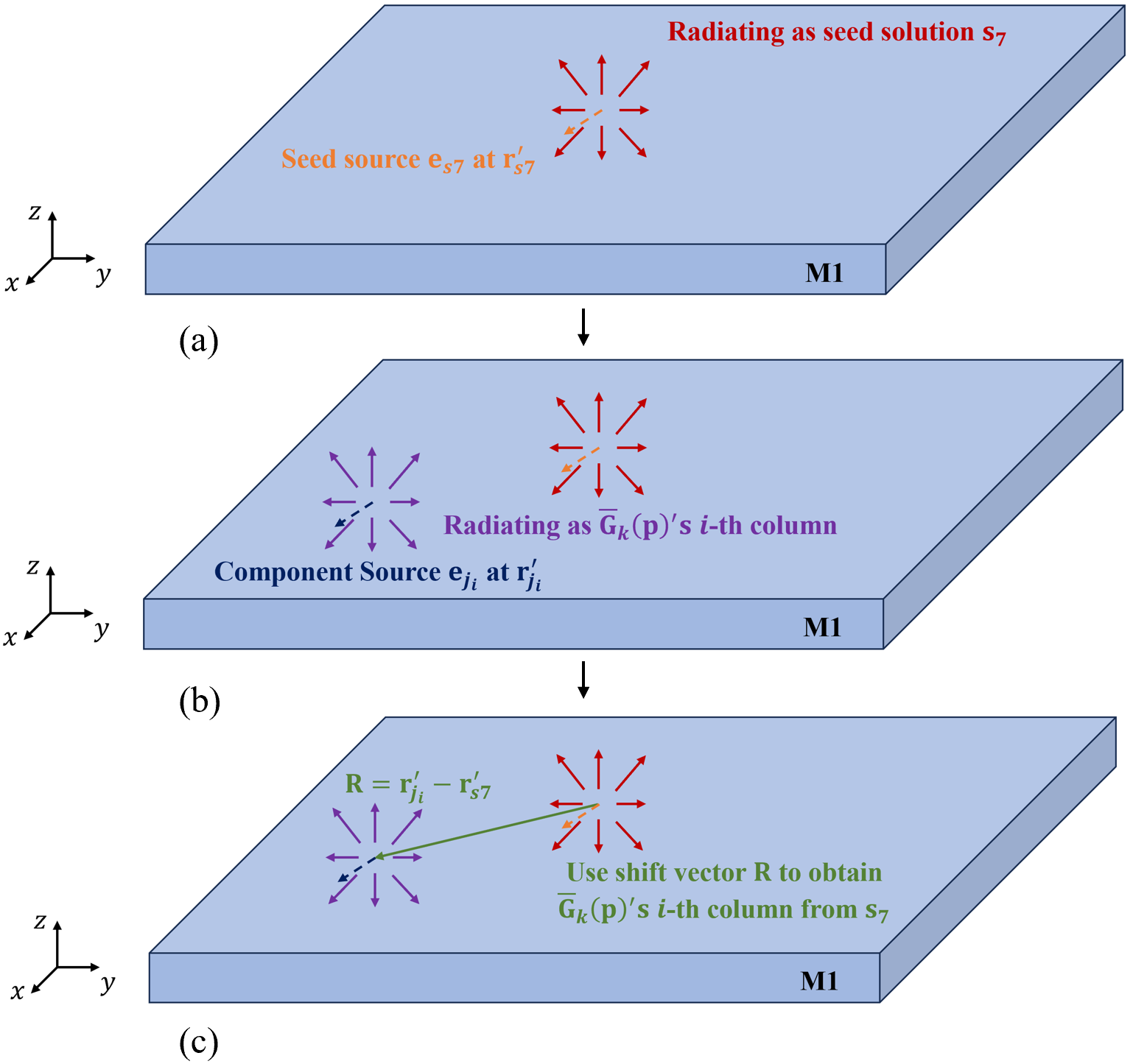}
\caption{Illustration of the proposed seed-and-shift method. (a) Seed excitation $\mathbf{e}_{s7}$ radiates as seed solution $\mathbf{s}_{7}$. (b) Component source $\mathbf{e}_{j_i}$ radiates as $\overline{\mathbf{G}}_k(\mathbf{p})$'s $i$-th column. (c) Calculate shift vector $\mathbf{R}$ to obtain $\overline{\mathbf{G}}_k(\mathbf{p})$'s $i$-th column from $\mathbf{s}_{7}$. }
\label{fig:GFShiftScheme}
\vspace{-0.15in}
\end{figure}

To overcome this, we utilize a key physical property of the invariant background in RFICs. An RFIC is typically fabricated in a layered dielectric stack. Let the layer-growth direction be $z$. Each layer is homogeneous in the $x-y$ plane. Consequently, $\overline{\mathbf{Y}}_b^{-1}$ is a function of the distance vector between a source point and an observation point in the $x-y$ plane within each layer. Hence, the field solution obtained from one source point can be shifted to obtain that due to another source point in the same layer. We thus only need to compute a small set of seed solutions: two horizontal seed solutions for $x$- and $y$-oriented unit vectors located on each $x-y$ surface where variant components exist, and one vertical seed solution generated with a $z$-oriented unit source vector for each $z$-layer where the variant components exist. Once the seed solutions are computed and stored, any of the $k$ columns of $\overline{\mathbf{G}}_k(\mathbf{p})$ can be generated almost instantaneously. The same technique applies to the computation of $\overline{\mathbf{Y}}_b^{-1}$ multiplied by an arbitrary vector, which is used in \eqref{equ:eb} for computing $\mathbf{e}_b$. As shown in Fig. \ref{fig:GFShiftScheme}, to obtain the $i$-th column corresponding to a component edge $j_i$, we match the edge to its corresponding seed excitation (e.g., $\mathbf{s}_7$ in the figure) and compute the $x-y$ plane shift vector $\mathbf{R}=\mathbf{r}_{j_i}^{\prime}-\mathbf{r}_{s}^{\prime}$. We then sample the matched seed solution $\mathbf{s}$ (e.g., $\mathbf{s}=\mathbf{s}_7$) using the shift vector to obtain 
\begin{equation}
\label{equ:seedSolShift}
\overline{\mathbf{G}}_{k, i\text{-th col}}(\mathbf{r})=\mathbf{s}(\mathbf{r}-\mathbf{R}).
\end{equation}

The aforementioned algorithm drastically reduces the computational cost. Instead of solving for $k$ right-hand sides, we only need to solve for a constant number of seed solutions in the order of the layer number. In addition, it eliminates the memory cost of storing $N \times k$ dense matrix $\overline{\mathbf{G}}_k(\mathbf{p})$. The coupling term $\overline{\mathbf{I}}_v^T \overline{\mathbf{G}}_k$ in (\ref{equ:MExpression}) can be built on-the-fly via seed-solution shifting, and the multiplication involving $\overline{\mathbf{Y}}_b^{-1} \overline{\mathbf{I}}_v(\mathbf{p})$ in (\ref{equ:WoodburyESol}) can be achieved by solving the pre-factorized system with the incoming vector expanded by $\overline{\mathbf{I}}_v(\mathbf{p})$ as the excitation. 
\section{Numerical Results} \label{sec:numiercalexam}
We validate the proposed method through two stages. First, we verify the accuracy and efficiency of the proposed acceleration technique described in Section \ref{sec:effientGF}. Second, we apply the complete algorithm to a large-scale design space exploration of a complex three-transformer RFIC system. All simulations are performed on a node equipped with an NVIDIA A100 80GB GPU. The invariant background is chosen as a 9-layer stack adapted from the GlobalFoundries 22FDX process, including embedded ground planes.

\subsection{Validation of the Efficient Computation of Numerical Green's Function} 
We define a domain with $N=632,109$ unknowns and a random set of component edges having $k=15,202$.  We test the computational efficiency of assembling the $k \times k$ coupling matrix $\overline{\mathbf{I}}_v^T \overline{\mathbf{G}}_k$. The brute-force method, which involves computing all $k=15,202$ columns of $\overline{\mathbf{G}}_k$, took 170.80 seconds. In contrast, the proposed seed-and-shift approach, which only computes $N_{\text{seeds}}=10$ seed solutions and then builds the matrix on-the-fly, took only 1.49 seconds. This demonstrates a 114.63$\times$ speedup. The relative difference between the $k \times k$ matrices generated by the proposed method in comparison with a brute-force computation of $\overline{\mathbf{G}}_k$ using the Frobenius norm is shown to be $1.33 \times 10^{-10}$, indicating excellent accuracy.

\subsection{Design Exploration of an RFIC Composed of Three-Transformers}
\begin{figure}[!t]
\centering
\includegraphics[width=0.8\columnwidth]{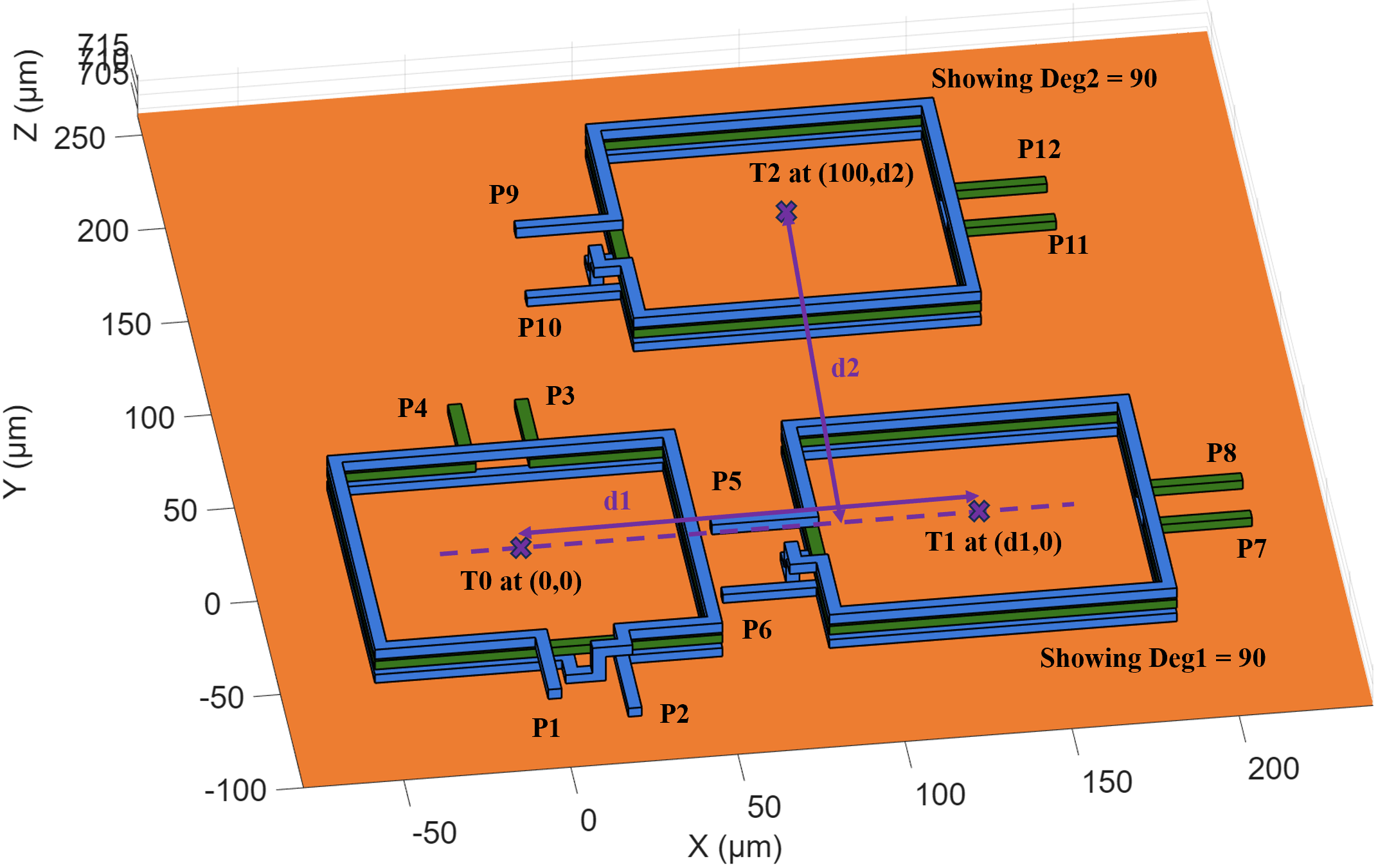}
\caption{3-D illustration of the three-transformer system.}
\label{fig:threeTrans}
\vspace{-0.15in}
\end{figure}

We apply the proposed algorithm to the design of an RFIC composed of three transformers shown in Fig. \ref{fig:threeTrans}. The unit transformer design is adapted from \cite{Hsu:2012:OnChipTransformer}. Each transformer unit is a four-port device, resulting in a 12-port system. The domain is discretized with a uniform 4 $\mu\mathrm{m}$ cell size in $x$ and $y$ directions, and one cell per layer in the $z$ direction, resulting in $N=745,589$ total unknowns. A 4-parameter sweep is defined, where Transformer T0 is fixed at the origin, and we vary T0-T1 center distance $d_1$, T0-T2 $y$-direction distance $d_2$, and the orientations of T1 (Deg1) and T2 (Deg2). This sweep constitutes a total of $17 \times 8 \times 2 \times 2 = 544$ unique design variations. The total number of component-related edges is $k=29,023$.

The simulation is run at 54 GHz. First, the full-domain responses for all 544 designs (each with 12 port excitations) are computed. The brute-force method, requiring a new $N \times N$ solve for each design, took 4792.64 s. The proposed algorithm's total time was 128.91 s. This is composed of a 9.23 s one-time cost for factorizing $\overline{\mathbf{Y}}_b$ and computing the seed solution, and 119.68 s to solve the $k \times k$ reduced system and superpose results for all 544 designs. This achieves a 37.2$\times$ total time reduction and a 40$\times$ speedup (4792.64 s / 119.68 s) at the design exploration step. Accuracy is also maintained; the F-norm relative difference for the $\overline{\mathbf{I}}_v^T \overline{\mathbf{G}}_k$ matrix was $2.42 \times 10^{-11}$, and the relative differences for the 12$\times$12 Z-parameter matrices across all 544 designs were all less than $1 \times 10^{-8}$.

\begin{figure}[!t]
\centering
\includegraphics[width=0.7\columnwidth]{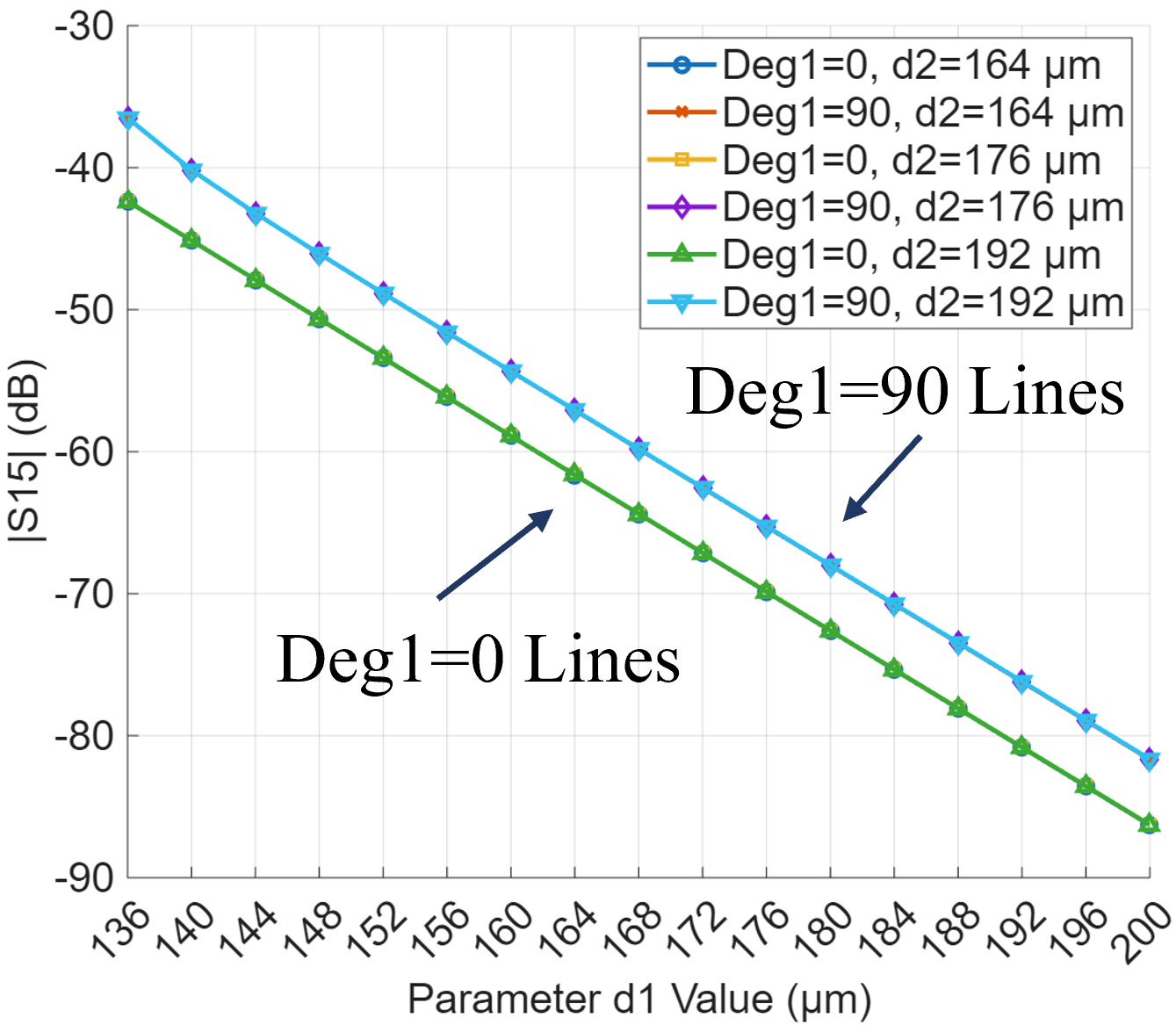}
\caption{Representative crosstalk result: Magnitude of $S_{15}$ at 54 GHz as a function of the center-to-center distance $d_1$.}
\label{fig:transformerPlaceSPara}
\vspace{-0.15in}
\end{figure}

Fig. \ref{fig:transformerPlaceSPara} shows a representative crosstalk result: $|S_{15}|$ versus the center-to-center distance $d_1$. The results illustrate the coupling trend across the parametric sweep, demonstrating the capability of the algorithm for rapid design exploration.

\section{Conclusion} \label{sec:conclusion}
In this paper, a first-principles accurate PDE-based algorithm is developed that algebraically separates the total electromagnetic response into an invariant background solution and a design-dependent correction from variant components. This formulation reduces the computational cost of design exploration to a small $k \times k$ system involving unknowns in the variant components only. An accurate and efficient method for fusing the component model has also been developed. The field solution in the invariant background for any source constitutes a numerical Green's function of the background. A seed-and-shift technique is developed to compute the Green's function efficiently by exploiting the $x-y$ spatial invariance in RFICs. The proposed algorithm was successfully applied to large-scale, multi-parameter sweeps of complex RFIC designs. Numerical experiments have demonstrated excellent accuracy while achieving significant computational speedups for large-scale design space exploration.



\end{document}